# Mapping mobility patterns to public spaces in a medium-sized city using geolocated tweets

*María Henar Salas-Olmedo*

Departamento de Geografía Humana, Facultad de Geografía e Historia, Universidad Complutense Madrid, Calle Profesor Aranguren, s/n, Ciudad Universitaria, 28040, Madrid, Spain

Email: msalas01@ucm.es

*Carolina Rojas Quezada*

Departamento de Geografía, Facultad de Arquitectura, Urbanismo y Geografía, Universidad de Concepción, Centro de Desarrollo Urbano Sustentable, CEDEUS, Calle Victoria s/n, Barrio Universitario, Concepción Región del Bíobío, Chile

Email: crojasq@udec.cl

Keywords: mobility; public space; big data; Twitter; medium-sized city; Chile

**Abstract:** This research evidences the usefulness of open big data to map mobility patterns in a medium-sized city. Motivated by the novel analysis that big data allow worldwide and in large metropolitan areas, we developed a methodology aiming to complement origin-destination surveys with à la carte spatial boundaries and updated data at a minimum cost. This paper validates the use of Twitter data to map the impact of public spaces on the different parts of the metropolitan area of Concepción, Chile. Results have been validated by local experts and evidence the main mobility patterns towards spaces of social interaction like malls, leisure areas, parks and so on. The map represents the mobility patterns from census districts to different categories of public spaces with schematic lines at the metropolitan scale and it is centred in the city of Concepción (Chile) and its surroundings (~10 kilometres).

1. Introduction

The study of social networks has long been a field of interest from different perspectives. Transport oriented studies are a major research line as they aim to uncover mobility patterns in order to improve transport and urban planning. There is however another raising field of interest related to the use of public spaces, with the focus on the shared spaces. The aim is to help reducing social exclusion by promoting the shared use of public spaces as opposed to the existence of segregated places.

In this research, we explore the utility of big open data through the Twitter platform for small to medium sized cities, which have been paid much less attention and still host almost 50 per cent of the urban population (United Nations 2014). We use the city of Concepción (Chile) as a study area to analyse mobility patterns to public spaces. Concepción is a medium-sized city with over 200,000 inhabitants and strong spatial relationship with Talcahuano. Both cities are the articulators of Concepción metropolitan area with nearly 1 million inhabitants.

Twitter data is publicly available, even at no cost if one uses the Streaming API. For this reason, Twitter has become a popular data source and is present in millions of research documents. However, most of them do not make use of its geographical dimension (Leetaru et al. 2013), or do it at a worldwide or regional scale (Hawelka et al. 2014; Sobolevsky et al. 2015; Li, Goodchild, and Xu 2013; Liu et al. 2015). The utility of Twitter data to obtain regular mobility data in large urban areas was demonstrated by Huang and Wong (2015) in their case study in Washington DC. They were able to obtain space-time paths from Twitter geolocated data similar to traditional travel diaries. We aim to extract similar material for our case study area in order to obtain information about the use of public spaces as opportunities for spatio-temporal interaction.

Our aim is to provide complementary and easily updatable information of mobility patterns to current origin-destination surveys that take place every 10 years, especially considering the last one in Concepción was done in 1999. Also, the public space is relevant because many social activities outside homes occur in public spaces (Rojas et al. 2015), even park and plazas play an important role for people (Villagra-Islas and Alves 2016). The challenge is to extract useful information from a smaller dataset than the ones available for large cities. In this paper we use mobility patterns to analyse the use of the public space by people residing in different parts of the city.

2. **Data and methods**

In order to fulfil the objective of improving spatio-temporal resolution and reducing data collection costs, we made use of geolocalized Twitter data. This is freely available on a real-time basis; thus an effort was made to download all the geolocalized tweets trough the streaming API in order to build a proper database. We used Python language to screen and save the tweets within a certain bounding box, and then we transformed it into a point layer for use in a Geographic Information System.

This research was made with the geolocalized tweets published in the study area (Concepción metropolitan area) between 1st January and 31 March 2016. Data treatment included the removal of identical tweets referring to the same emergency phenomena and at the same location, since this information is not relevant for the purposes of this research, i.e. identifying the use of public space.

In addition, user accounts with more than 250 tweets over the whole period (> 2.7 tweets per day on average) were checked in order to remove those not corresponding to individuals (i.e. Twitter accounts devoted to disseminate news or emergency issues). This led to the removal of 17 of the 26 top active users in the entire metropolitan area. Table 1 below shows the main figures of our sample data referring to the regional, metropolitan and central core of Concepción.

Table 1. Main figures of downloaded geo-located tweets in Concepción, Chile

|  | Total tweets | Valid tweets | | Users | Users that moved* | |
| --- | --- | --- | --- | --- | --- | --- |
|  | No. | No. | % | No. | No. | % |
| **Regional area** | 52536 | 37838 | 72.02 | 4113 | 2258 | 54.90 |
| **Metropolitan area** | 52345 | 37708 | 72.04 | 4101 | 2255 | 54.99 |
| **Central area** | 21568 | 17422 | 80.78 | 2494 | 1695 | 67.96 |

* Users that tweeted more than once and from different locations.

Source: own elaboration from data obtained from Twitter Public API between 01 January 2016 and 31 March 2016, local time (GMT -3).

The generation of mobility maps implies the need to set Twitter users' place of residence, which was done assuming they live in the district were they tweeted the most between 22:00 and 07:59 on regular weekdays (Monday to Thursday).

Our methodology includes an exploratory analysis of the spatial autocorrelation between official resident-based data and our estimated place of residence prior to the flow maps representing mobility patterns. This way we aim to confirm the null hypothesis that the number of users of public spaces from a particular place is not directly proportional to that place population volume and distance, and that this relationship varies across the space.

In particular, we used global and local bivariate Moran's spatial autocorrelation Index, which indicates whether there is a strong or weak relationship between high (and low) population volume and Twitter users at the district level. This index was developed by Anselin et al. (2010) and was computed using their GeoDa software.

The mobility analysis was made in base to those users that tweeted at night on working days at any part of the metropolitan area, and during the day and weekends within the most popular public spaces according with local researchers. Data treatment and map production was done with the commercial software ArcGIS 10.3.1 and Simantel's Flow Map Generator (Simantel 2012) toolbox.

We followed Shelton (2016) and proceed to normalize our data to a tweet usage baseline. The mobility analysis was therefore performed with 'raw' data (i.e. number of Twitter users moving from one district to a particular public space) and with normalized data (i.e. the proportion of Twitter users from one district to a particular public space). The comparison between the former (number of people moving) and the latter (the impact of each public space on each district) allows the identification of a potential bias in using Twitter as a proxy for individual mobility.

3. **Mapping mobility patterns**

We analysed the mobility patterns between the estimated district of residence and a selection of public spaces. These public spaces were selected and mapped according to local experts and represent the most popular places in the metropolitan area in six categories, i.e. Central Business District (CBD), Malls, Leisure, University Campus, Transport and Parks (Figure 1).

Figure 1. Categories of public spaces

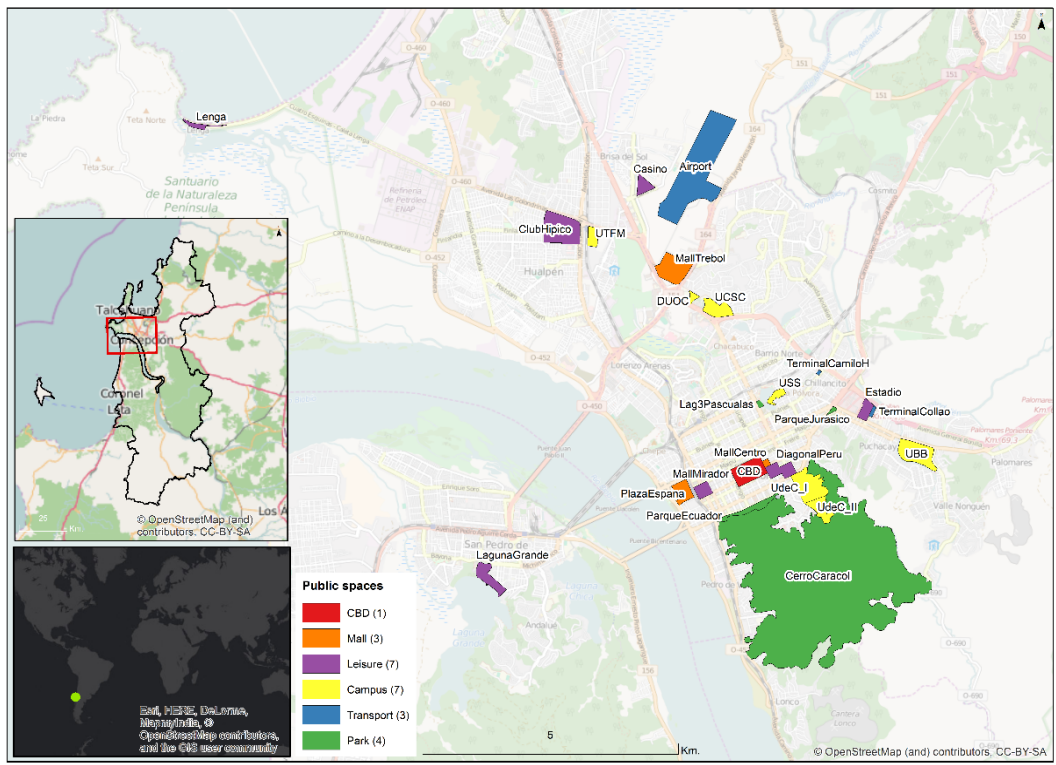

Source: own elaboration from OpenStreetMap. Colour symbols: ColorBrewer.org.

The exploratory analysis of spatial autocorrelation evidences a weak relationship between resident population (according to district delimitations of the latest Census, 2002) and the estimated place of residence of Twitter users that visited the selection of public spaces (Figure 2, left). This indicates that, globally, the resulting flows between place of residence and public spaces are not influenced by the population volume in each census district. Figure 2's right panel shows the spatial pattern of this relationship, thus adding useful insights. Districts in red are those with high volume of both variables, while light blue indicates low values of population surrounded by high volume of Twitter users that visited the public spaces. These two categories are statistically significant in the northern and central part of Concepción metropolitan area. On the contrary, the southern part, which is a rural area, registers statistically significant low values of both variables, with the exception of Santa Juana district (high population, low Twitter users).

Figure 2. Twitter users that visited public spaces by district of residence and resident population scatterplot (left) and local Moran's I (right)

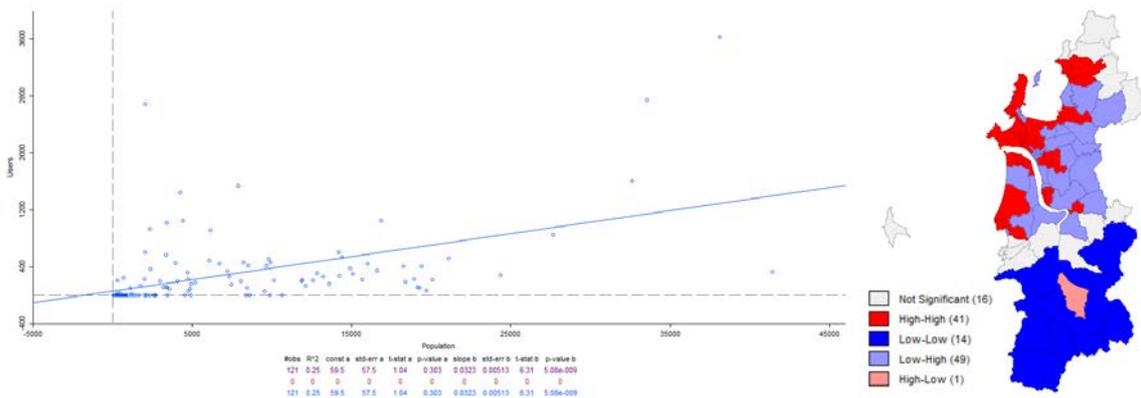

Source: own elaboration from data obtained from Twitter Public API between 01 January 2016 and 31 March 2016, local time (GMT -3) and 2002 Census (Instituto Nacional de Estadística de Chile INE).

We then compared the results of non-normalized vs. normalized data in order to understand the mobility patterns to public spaces (Figure 3, top vs. bottom). These maps show the number of Twitter users in each district that visited each public space (top, non-normalized data) vs. the impact of each public space on residential areas by district (bottom, normalized data). Line thickness grows proportionally with the number of users/size of the impact. Lines start at the furthest district in a given direction and accumulate the flow from nearby districts which lines merge together (like a hydrological basin). The scale growth is constant in all maps, thus making them easily comparable. We produced generalized lines that are topological and quantitative accurate, and that maximize map simplicity and readability.

CBD's impact is much larger than any other public space, thus deserving a prominent place. The rest of the public spaces are grouped in maps by categories. Colours representing individual public spaces in each map have been carefully chosen from ColorBrewer qualitative ramps (Brewer 2002) in order to ensure equitable visibility to each one.

As expected, the CBD is the main attraction place with users from all districts and a greater impact on the north-west and south part of the metropolitan area (flows from Talcahuano, Penco, San Pedro de la Paz and Coronel). Normalized flows indicate that Trebol mall (the largest in size and shops), the furthest away from the city centre and the easiest to reach by car, has the largest impact on most districts, including the city centre. On the contrary, Centro mall shows a sharp decrease in its impact outside the inner city boundaries (i.e. walkable and with higher transit density), and the Mirador mall extends most of its impact along the northwest-southeast axis, which is coincident with the railway and bus network. University and college campus also concentrate a high proportion of Twitter visitors with differences across the metropolitan area. Universidad San Sebastián (USS), on the northern part of the city centre, gets the highest impact from its surroundings and, to a lesser extent, from the western area. In the second position, University of Concepción (UdeC) has a large impact that spreads in all directions, whereas DUOC and UBB's impact decreases sharper with distance.

Leisure areas not related to retail activities but to entertainment and gastronomy receive fewer visitors. In some cases, the impact is clearly concentrated on some specific parts of the metropolitan area. For example, Diagonal-Perú, an outdoor bar and restaurant area in the city centre, has a higher impact on the northern districts, whereas Casino, a private entertainment area in the northern outskirts, impacts especially in the districts located on the high-income residential area north of the city centre and on the southwest part of the metropolitan area. Both areas also receive a larger impact from the airport than other districts. Other places, like Lenga, Plaza España or Laguna Grande have an impact that decreases with distance in a consistent way. In the case of parks, it is clear that Parque Ecuador, the one with a clear urban character, is the only one with a hinterland covering the whole metropolitan area. Similarly, Terminal Collao bus station has an extended impact which is more intense towards the north and west.

Figure 3. Twitter users' flows from estimated resident district to public spaces

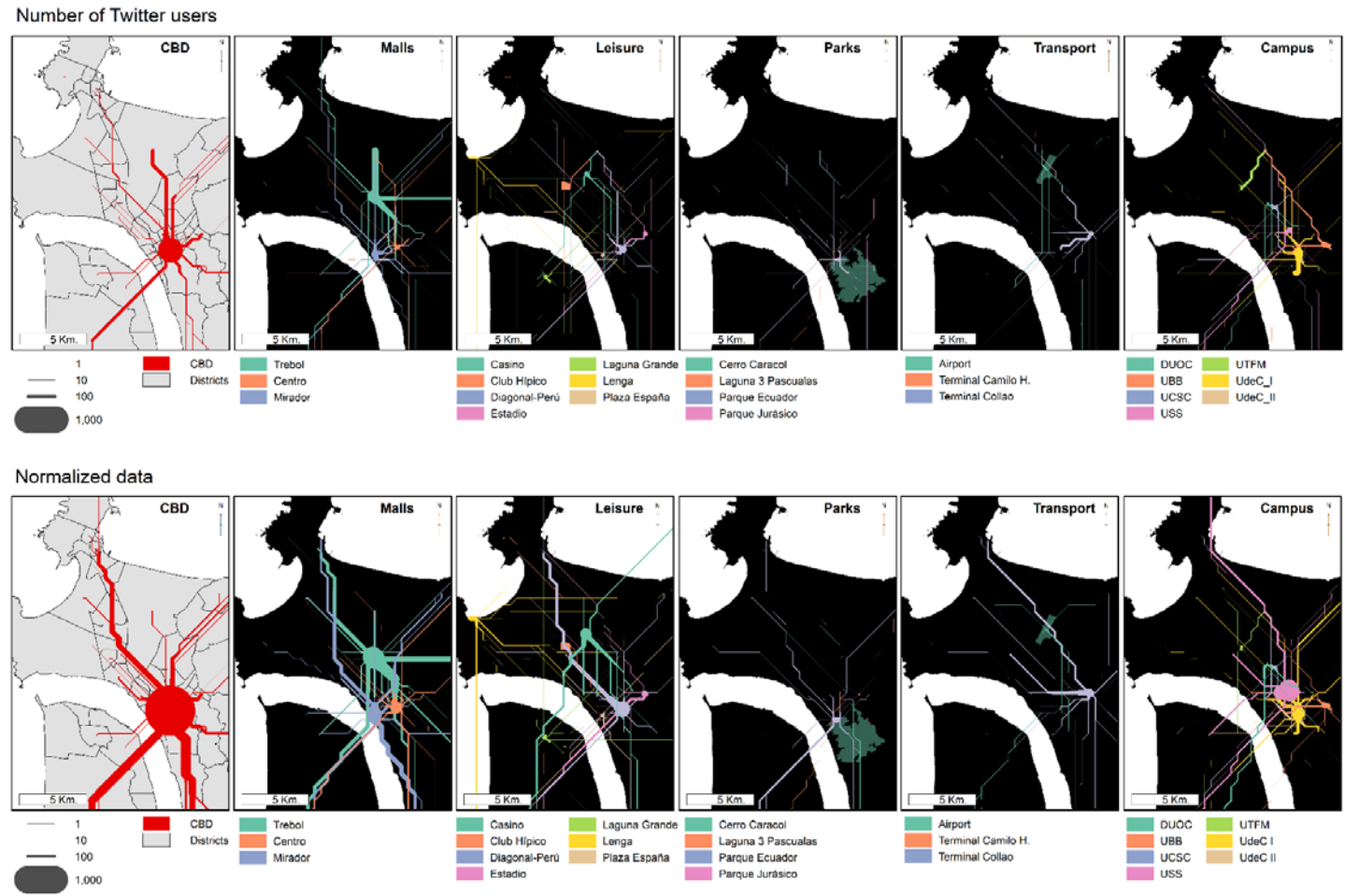

Source: own elaboration from data obtained from Twitter Public API between 01 January 2016 and 31 March 2016, local time (GMT -3). Colours: ColorBrewer.org.

## 4. Conclusions

In this research, we used a 3-month sample data to investigate the potential of geolocated tweets to map mobility patterns in a medium-sized city. We were interested in mapping the mobility patterns associated with different types of public spaces in order to unveil potential areas of social exclusion: public spaces which are less visited by people living relatively near, which prefer further spaces.

Our methodology is based on Geographic Information Systems (GIS) and it includes an exploratory bivariate spatial autocorrelation analysis to help the interpretation of the results as well as flow maps to compare normalized vs. non-normalized data.

Exploratory analysis indicates that the spatial distribution of public space users differs from population density, with a significant gap between the northern and central metropolitan area, and the southern part in terms of Twitter usage.

Flow maps between the estimated district of residence and a selection of public spaces were then computed and represented in the form of schematic lines representing accumulative flows (raw data) or impact (normalized data) towards each public space. According to local experts, maps showing normalized data (i.e. the impact of each public space over each district of residence) represent close-to-reality mobility patterns. One clear example is the case of the malls, which impact's spatial pattern is similar to the transport network of the mode that provides higher accessibility in each case. This results of great value to complement current mobility surveys and other data sources.

Further research includes the validation of a 6-month sample with the results of a recent origin-destination survey and prospections to produce finer time and spatial resolution.

**Software**

We used several technologies in each step. First, data collection was based on the Tweepy Python library (Roesslein 2009), which was modified to fit our needs. Data treatment to convert geolocated tweets to a GIS point layer was done with Stata 12 and ESRI's ArcGIS Desktop 10.3.1. The spatial autocorrelation analysis was computed with GeoDa and the lines representing the flows were generated with Simantel's Flow Map Generator toolbox for ArcGIS. Colour selection was based on ColorBrewer qualitative palettes. The final composition was made on ESRI's ArcMap 10.3.1.

**Acknowledgements**

The authors would like to thank Santander Universidades for financing a three-month mobility grant to Universidad de Concepción (Chile) and the Ministerio de Economía y Competitividad of Spain and Universidad Complutense Madrid for the funding provided for a post-doctoral fellowship (FPDI-2013-17001). This research is also supported by Centro de Desarrollo Urbano Sustentable CEDEUS (FONDAP CONICYT 15110020), research project TRA2015-65283-R (Ministerio de Economía y Competitividad of Spain), and programme S2015/HUM-3427 (Comunidad de Madrid and European Structural Funds). Special thanks must be given to Gaspar Cascallana, Guillermo Dorado and Borja Moya-Gómez for their valuable help to access the data.

**References**

Anselin, Luc, Ibnu Syabri, and Youngihn Kho. 2010. "GeoDa: An Introduction to Spatial Data Analysis." In *Handbook of Applied Spatial Analysis: Software Tools, Methods and Applications*, edited by M Manfred Fischer and Arthur Getis, 73–89. Berlin, Heidelberg: Springer Berlin Heidelberg. doi:10.1007/978-3-642-03647-7_5.

Brewer, Cynthia. 2002. "ColorBrewer." http://www.colorbrewer.org.

Hawelka, Bartosz, Izabela Sitko, Euro Beinat, Stanislav Sobolevsky, Pavlos Kazakopoulos, and Carlo Ratti. 2014. "Geo-Located Twitter as Proxy for Global Mobility Patterns." *Cartography and Geographic Information Science* 41 (3). Taylor & Francis: 260–71. doi:10.1080/15230406.2014.890072.

Huang, Qunying, and David W. S. Wong. 2015. "Modeling and Visualizing Regular Human Mobility Patterns with Uncertainty: An Example Using Twitter Data." *Annals of the Association of American Geographers* 105 (6). Routledge: 1179–97. doi:10.1080/00045608.2015.1081120.

Leetaru, Kalev, Shaowen Wang, Anand Padmanabhan, and Eric Shook. 2013. "Mapping the Global Twitter Heartbeat: The Geography of Twitter." *First Monday* 18 (5). doi:10.5210/fm.v18i5.4366.

Li, Linna, Michael F. Goodchild, and Bo Xu. 2013. "Spatial, Temporal, and Socioeconomic Patterns in the Use of Twitter and Flickr." *Cartography and Geographic Information Science* 40 (2). Taylor & Francis: 61–77. doi:10.1080/15230406.2013.777139.

Liu, J, K Zhao, S Khan, M Cameron, and R Jurdak. 2015. "Multi-Scale Population and Mobility Estimation with Geo-Tagged Tweets." In *Data Engineering Workshops (ICDEW), 2015 31st IEEE International Conference on*, 83–86. doi:10.1109/ICDEW.2015.7129551.

Roesslein, Joshua. 2009. "Tweepy. An Easy-to-Use Python Library for Accessing the Twitter API." http://www.tweepy.org/.

Rojas, Carolina Alejandra, Juan Antonio Carrasco, Fabian Pérez, Gerson Araneda, and Claudia

Lima. 2015. "Análisis de Redes Sociales." In *XV Conferencia Iberoamericana de Sistemas de Información Geográfica*. Valparaíso.

Simantel, Brad. 2012. "Simantel, B. (2012). Flow Map Generator. ArcGIS Toolbox." http://www.arcgis.com/home/item.html?id=3bb350838ee044979d1f583b116cb8c1.

Sobolevsky, S, I Bojic, A Belyi, I Sitko, B Hawelka, J M Arias, and C Ratti. 2015. "Scaling of City Attractiveness for Foreign Visitors through Big Data of Human Economical and Social Media Activity." In *2015 IEEE International Congress on Big Data*, 600–607. doi:10.1109/BigDataCongress.2015.92.

United Nations. 2014. *World Urbanization Prospects: The 2014 Revision, Highlights (ST/ESA/SER.A/352). New York, United.* doi:10.4054/DemRes.2005.12.9.

Villagra-Islas, Paula, and Susana Alves. 2016. "Open Space and Their Attributes, Uses and Restorative Qualities in an Earthquake Emergency Scenario: The Case of Concepción, Chile." *Urban Forestry & Urban Greening* 19: 56–67. doi:10.1016/j.ufug.2016.06.017.